\documentclass[pre,reprint,twocolumn,showpacs,preprintnumbers,amsmath,amssymb]{revtex4-1}

\usepackage{graphicx}
\usepackage{dcolumn}
\usepackage{bm}
\usepackage{tikz,pgf}

\begin{document}

\preprint{APS/123-QED}

\title{Magnetohydrodynamic Turbulent Cascade of Coronal Loop Magnetic Fields}

\author{A. F. Rappazzo$^{1,2,3}$}
\author{M. Velli$^{2}$}
\affiliation{%
$^1$Instituto de Astrof\'{\i}sica de Canarias, I-38200 La Laguna, Tenerife, Spain\\
$^2$Jet Propulsion Laboratory, California Institute of Technology, Pasadena, California 91109, USA\\
$^3$Bartol Research Institute, Department of Physics and Astronomy, University of Delaware, Delaware 19716, USA}

\begin{abstract}
The Parker model for coronal heating is investigated 
through a  high resolution simulation.
An inertial range is resolved where fluctuating magnetic energy 
$E_{_M}(k_{\perp}) \propto k_{\perp}^{-2.7}$ exceeds
kinetic energy $E_{_K}(k_{\perp}) \propto k_{\perp}^{-0.6}$.
Increments scale as $\delta b_{_{\!\ell}} \simeq  \ell^{-0.85}$ and
$\delta u_{_{\!\ell}} \simeq  \ell^{+0.2}$ with velocity increasing at small scales,
indicating that magnetic reconnection plays a prime role in this turbulent system.
We show that spectral energy transport is akin to standard magnetohydrodynamic (MHD) turbulence
even for a system of reconnecting current sheets sustained by the boundary.
In this new MHD turbulent cascade, kinetic energy flows are negligible while
cross-field flows are enhanced, and through a series of ``reflections'' between the two fields, 
cascade more than half of the total spectral energy flow.
\end{abstract}


\maketitle

The heating of solar and stellar atmospheres is an outstanding astrophysical
problem~\cite{re10}.
The solar corona has temperatures ($\gtrsim 10^6\, K$) up to
three orders of magnitude higher than the underlying photospheric and chromospheric
layers, sustained by an energy flux of about 
$10^7\, erg\, cm^{-2}\, s^{-1}$~\cite{Withbroe:1977p363}.
 
Most of the coronal x-ray and extreme ultraviolet (EUV) radiation is emitted in 
loops, bright structures threaded by a strong axial magnetic 
field connecting photospheric regions of opposite polarity.
In the scenario proposed by Parker~\cite{Parker:1972p1352},
magnetic field lines are braided by convective photospheric motions 
that shuffle their footpoints, leading to the ``spontaneous'' development 
of small-scale current  sheets where the  plasma is  heated.

It has long been proposed~\cite{Einaudi:1996p729} that this scenario
can be regarded as a magnetohydrodynamic (MHD) turbulence problem as photospheric motions 
stir the magnetic field lines' footpoints, and these motions are transmitted 
inside by the field-line tension stirring in this way (anisotropically) the whole 
plasma akin to a body force.
Simulations have indeed revealed that the system exhibits many properties 
of an authentic MHD turbulent system, including the formation of field-aligned current sheets,
power-law spectra for the energies, and power-law distributions for energy 
release, peak dissipation and duration of dissipative 
events~\cite{Dmitruk:1997p1563,*Einaudi:1999p707,*Dmitruk:2003p499,%
Rappazzo:2007p331,*Rappazzo:2008p344,*Rappazzo:2010p4099}.
Furthermore in recent papers~\cite{Rappazzo:2007p331,*Rappazzo:2008p344} we have 
developed a phenomenological scaling model for this turbulent cascade 
where at the large scales nonlinearity is weak (i.e., depleted similarly 
to~\cite{Galtier:2000p4461}) and at the small scales strong~\cite{Goldreich:1995p1640}.

However  this new, line-tied, turbulent regime 
is quite distinct from the classical MHD turbulence system where energies  are in 
approximate equipartition,  as here fluctuating 
magnetic energy dominates over kinetic energy throughout the inertial range.

Certainly, this system can also be seen as a set of 
reconnecting current sheets sustained by the boundaries, 
and a large fraction of the kinetic energy might be contributed by the magnetic
field through reconnection itself rather than from cascading large-scale kinetic energy.

It is therefore crucial to understand if turbulence is an appropriate framework to 
model this problem, namely, can a set of reconnecting current sheets
be described in terms of turbulence? The influence of turbulence on 
magnetic reconnection is an active research topic 
(e.g., see~\cite{Lazarian:1999p1657} for a model), 
but are energy fluxes in a system where magnetic reconnection 
plays a prime role similar to those of MHD turbulence? 
We present here a novel investigation of the energy 
flows between different scales  and fields in Parker's model in 
order to determine how different the spectral fluxes are 
and what properties they share with the standard MHD turbulence  
case~\cite{Dar:2001p2747,Alexakis:2005p2633,Aluie:2010p4029}.

A coronal loop is modeled in Cartesian geometry as a plasma with uniform  density 
$\rho_0$ embedded in a strong axial magnetic field $B_0$ directed  along $z$ 
(see~\cite{Rappazzo:2008p344} for a more detailed description of  the model and numerical 
code). Magnetic field lines are line tied at the top and bottom plates where a large-scale 
velocity field is imposed. In the perpendicular direction ($x$-$y$) periodic boundary conditions are 
used. The dynamics are modeled with the (non-dimensional) equations of reduced 
magnetohydrodynamics 
(RMHD)~\cite{Kadomtsev:1974p283,*Strauss:1976p1438}:\begin{eqnarray}
&&{\displaystyle \partial_t \mathbf{u}} 
+ \left( \mathbf{u} \cdot\! \mathbf{\nabla_{_{\!\!\perp}}}\!  \right)   \mathbf{u}
= - \mathbf{\nabla_{_{\!\!\perp}}}\! P
+   \left(  \mathbf{b} \cdot\! \mathbf{\nabla_{_{\!\!\perp}}}\!\right)  \mathbf{b} 
+ c_{_{\!A}} \partial_z \mathbf{b} + 
\frac{1}{R} \mathbf{\nabla_{_{\!\!\perp}}}^{\!\!\!2} \mathbf{u},  \nonumber \\
&&\partial_t \mathbf{b} + \left( \mathbf{u} \cdot\! \mathbf{\nabla_{_{\!\!\perp}}}\!\right)  \mathbf{b} 
= \left( \mathbf{b} \cdot\! \mathbf{\nabla_{_{\!\!\perp}}}\!\right)  \mathbf{u} 
+ c_{_{\!A}}  \partial_z \mathbf{u}
+ \frac{1}{R} \mathbf{\nabla_{_{\!\!\perp}}}^{\!\!\!2}
\mathbf{b}, \label{eq:adim} 
\end{eqnarray}
with $\mathbf{\nabla_{_{\!\!\perp}}}\!\! \cdot \mathbf{u} = 
\mathbf{\nabla_{_{\!\!\perp}}}\! \!\cdot \mathbf{b} = 0$.
Here gradient and Laplacian operators have only orthogonal ($x$-$y$) components as do 
velocity and magnetic field vectors ($u_z$=$b_z$=0), while $P$ is the total (plasma plus 
magnetic) pressure. $c_{_{\!A}}$ is the ratio between the Alfv\'en velocity of the axial field 
($B_0/\sqrt{4\pi\rho_0}$) and the rms of photospheric velocity ($1\, km\, s^{-1}$) and $R$ 
is the Reynolds number.
In the simulation presented here, $c_{_{\!A}}=400$, $R=2000$, and the domain spans
$0 \le x, y \le \ell$, $0 \le z \le L$, with $\ell=1, L=10$ and 
$1024^2 \times 512$ grid points.
Given the orthogonal Fourier transform
$\mathbf{u} (\mathbf{x}_{_{\!\perp}},z) = \sum_{\mathbf{k_{_{\!\perp}}}}
\hat{\mathbf{u}} (\mathbf{k_{_{\!\perp}}},z)\, e^{i\mathbf{k_{_{\!\perp}}}\!\! \cdot \mathbf{x}_{_{\!\perp}}}$ 
with $\mathbf{k_{_{\!\perp}}} = 2\pi\, \mathbf{n_{_{\!\perp}}}/ \ell$
and $\mathbf{n_{_{\!\perp}}}\! \in \mathbb{Z}^2$,
the shell-filtered field $\mathbf{u}_{_K}$ is  defined as~\cite{Alexakis:2005p2633}
\begin{equation}
\mathbf{u}_{_K} (\mathbf{x}_{_{\!\perp}},z) =  \sum_{\mathbf{n}_{_{\!\perp}} \in K}
\hat{\mathbf{u}} (\mathbf{k_{_{\!\perp}}},z)\, e^{i \mathbf{k_{_{\!\perp}}} \!\! \cdot \mathbf{x_{_{\!\perp}}}},
\label{eq:shflt}
\end{equation}
i.e.,  it has only components in the ``shell'' $K$ with wave numbers K-1$<\!|\mathbf{n_{_{\!\perp}}}|\! \le$K.
The boundary photospheric velocities at $z\!=\!0,L$ are given random amplitudes 
for all wave numbers 3$\le\! n_{_{\!\perp}}\! \le$4 and then normalized so that the 
rms value is $1/2$ \cite[see][]{Rappazzo:2008p344}. As a result the forcing  boundary 
velocity has only components in shells 3 and 4.

\begin{figure}
	\includegraphics[scale=.48]{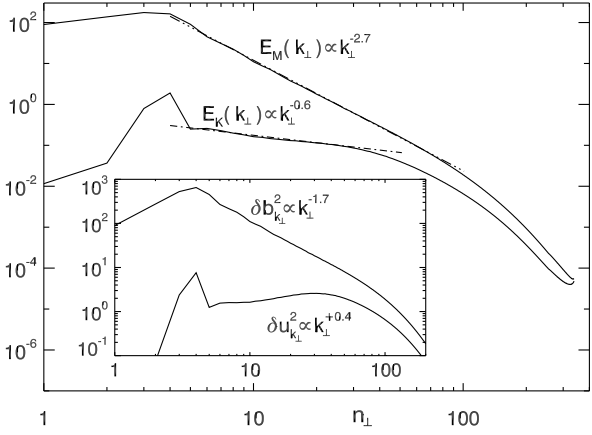}
	\caption{\label{fig1} Magnetic ($E_{_M}$) and kinetic ($E_{_K}$) energy 
	spectra as a function of the orthogonal wave number $n_{_{\!\perp}}$
	($k_{_{\!\perp}} = 2 \pi n_{_{\!\perp}}$). Inset: Increments, velocity 
	increasing at high wave numbers.}
\end{figure}

Since fields filtered in different shells are orthogonal,
and indicating the volume integrals with $\langle ... \rangle$,  
the equations for kinetic and magnetic energies 
$E_{_u}^{(K)} = \langle\, \mathbf{u}_{_K}^2 / 2\, \rangle$,
$E_{_b}^{(K)} = \langle\, \mathbf{b}_{_K}^2 / 2\, \rangle$
in shell $K$ follow from Eqs.~(\ref{eq:adim}):\vspace{-3pt}
\begin{eqnarray}
\partial_t E_{_u}^{(K)} & = &
\sum_{Q} \left[ \mathcal{T}_{_{uu}}^{(Q,K)} + \mathcal{T}_{_{bu}}^{(Q,K)} \right]
+ \frac{S^{(K)}}{2}
-\frac{\mathcal{D}_{_u}^{(K)}}{\mathrm{R}}\quad \label{eq:euk}\\
\partial_t E_{_b}^{(K)} & = &
 \sum_{Q} \left[ \mathcal{T}_{_{ub}}^{(Q,K)} + \mathcal{T}_{_{bb}}^{(Q,K)} \right]
+ \frac{S^{(K)}}{2}  - \frac{\mathcal{D}_{_b}^{(K)}}{\mathrm{R}}\label{eq:ebk}
\end{eqnarray}
These are obtained as for the three-periodic case~\cite{Alexakis:2005p2633},
except for terms $\propto \partial_z$ in Eqs.~(\ref{eq:adim}) that contribute
the integrated Poynting flux $S^{(K)}$ (the work done by photospheric motions on magnetic field lines' 
footpoints) entering the system at the boundaries in shell $K$:
\begin{equation}
S^{(K)}  = c_{_{\!A}} \left( \int_{z=L} \! \mathrm{d}a\ \mathbf{u}_{_K}\! \cdot \mathbf{b}
                                   -  \int_{z=0} \! \mathrm{d}a\ \mathbf{u}_{_K}\! \cdot \mathbf{b} \right).
\end{equation}
This does not cancel out along the nonperiodic axial direction $z$. 
As photospheric velocities have only components in shells 3 and 4
$S^{(K)}$ vanishes outside these two shells (\emph{the injection scale}).
In similar fashion the dissipative terms
$\mathcal{D}_{_u}^{(K)} = 
\langle \left| \mathbf{\nabla} \mathbf{u}_{_K} \right|^2 \rangle$,
$\mathcal{D}_{_b}^{( K )} = 
\langle \left| \mathbf{\nabla} \mathbf{b}_{_K} \right|^2 \rangle$
contribute only at dissipative scales with large $K$.

Between the injection and dissipative scales  only the following terms
contribute:
\begin{eqnarray}
&&\mathcal{T}_{_{uu}}^{(Q,K)} =  - \langle\, \mathbf{u}_{_K}\! \cdot 
\left( \mathbf{u} \cdot \mathbf{\nabla} \right) \mathbf{u}_{_Q} \, \rangle \label{eq:t9}\\[.1em]
&&\mathcal{T}_{_{bu}}^{(Q,K)} 
 =  \langle\, \mathbf{u}_{_K}\! \cdot 
\left( \mathbf{b} \cdot \mathbf{\nabla} \right) \mathbf{b}_{_Q} \, \rangle
+  \mathcal{A}_{_{bu}}^{( K )} \delta_{_{Q,K}} \label{eq:t11}\\[.1em]
&&\mathcal{A}_{_{bu}}^{( K )} = c_{_\mathcal{A}} 
\langle\, \mathbf{u}_{_K}\! \cdot \partial_z \mathbf{b}_{_K} \, \rangle - \frac{1}{2}\, S^{(K)}\\[.1em]
&&\mathcal{T}_{_{bb}}^{(Q,K)} =  - \langle\, \mathbf{b}_{_K}\! \cdot 
\left( \mathbf{u} \cdot \mathbf{\nabla} \right) \mathbf{b}_{_Q} \, \rangle \label{eq:t10} 
\end{eqnarray}
\begin{eqnarray}
&&\mathcal{T}_{_{ub}}^{(Q,K)} =  \langle\, \mathbf{b}_{_K}\! \cdot 
\left( \mathbf{b} \cdot \mathbf{\nabla} \right) \mathbf{u}_{_Q} \, \rangle
+  \mathcal{A}_{_{ub}}^{( K )} \delta_{_{Q,K}} \label{eq:t13}\\[.1em]
&&\mathcal{A}_{_{ub}}^{( K )} = c_{_\mathcal{A}} 
\langle\, \mathbf{b}_{_K}\! \cdot \partial_z \mathbf{u}_{_K} \, \rangle - \frac{1}{2}\, S^{(K)} \label{eq:t14}
\end{eqnarray}
These represent energy fluxes between fields in different shells. In fact given two 
fields $\mathbf{v}$ and $\mathbf{w}$ (either  velocity or magnetic fields)  the relation
$\mathcal{T}_{_{vw}}^{(Q,K)} = - \mathcal{T}_{_{wv}}^{(K,Q)}$
holds for the flux between shells $Q$ and $K$, 
and together with Eqs.~(\ref{eq:euk})-(\ref{eq:ebk}) define the fluxes 
\cite{Alexakis:2005p2633}. For example, \ $\mathcal{T}_{_{ub}}^{(Q,K)} $ represents 
conversion of kinetic energy in shell $Q$ to magnetic energy in shell $K$.

\begin{figure}
	\includegraphics[scale=.47]{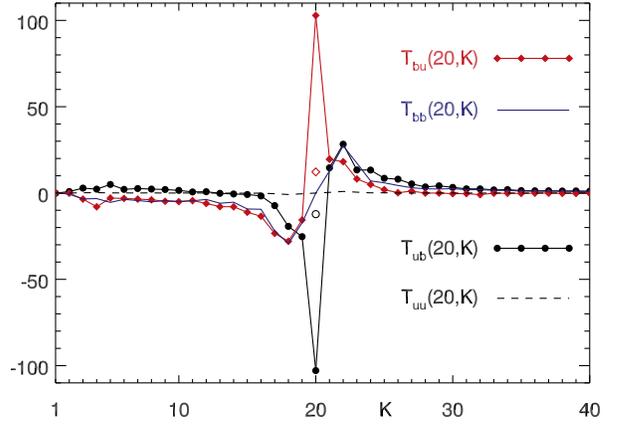}
	\caption{\label{fig2} (Color online) Spectral fluxes [Eqs.~(\ref{eq:t9}), (\ref{eq:t10}), (\ref{eq:t11}),
	 and (\ref{eq:t13})] showing incoming and outgoing energy transfers  between different
	 shells and fields to/from shell $Q$=20.}
\end{figure}

$\mathcal{A}_{_{ub}}^{( K )}$ is the flux due to the linear
terms $\propto \partial_z$ in Eqs.~(\ref{eq:adim}).
It does not transfer energy between different shells, but only
in the same shell between fields $\mathbf{u}$ and $\mathbf{b}$. 
Notice that it has been symmetrized  so that 
$\mathcal{A}_{_{ub}}^{(K)} = - \mathcal{A}_{_{bu}}^{(K)}$, 
as can be verified integrating by parts.

Equations~(\ref{eq:euk})-(\ref{eq:ebk}) show that at the injection scale 
(shells 3 and 4) the photospheric forcing supplies energy \emph{at the same 
rate} to both kinetic and magnetic energies, i.e., the forcing injects Alfv\'en
waves unlike standard forced MHD turbulence where a mechanical force injects only 
kinetic energy.

The simulation is started with vanishing orthogonal velocity and 
magnetic fields ($\mathbf{u} = \mathbf{b} = 0$) and a uniform axial
magnetic field $\mathbf{B} = B_0\, \mathbf{\hat{e}_z}$ inside the 
computational box.
As shown in our previous works~\cite{Rappazzo:2008p344} the 
constant forcing velocity at the boundary advects magnetic 
field lines generating a perpendicular component $\mathbf{b}$.
This initially grows linearly in time before saturating nonlinearities develop, 
and kinetic and magnetic energies then fluctuate around a mean value, 
with magnetic field fluctuations dominating: 
$E_{_M}/E_{_K}\! \sim 87$ for the simulation presented here.

The magnetic and kinetic energy imbalance is reflected in the energy 
spectra shown in Fig.~\ref{fig1}.
Both spectra are peaked at the injection wave numbers 3 and 4, but
beyond $n_{_{\!\perp}}\!$=\,5 an inertial range is resolved where
both spectra exhibit a power-law behavior with a steep index
for the magnetic energy ($-2.7$) and a flatter one for the kinetic 
energy ($-0.6$). Since increments are obtained from band-integrated 
spectra [e.g., $\delta b^2(k_{_{\!\perp}}\!) \simeq  k_{_{\!\perp}} E_M(k_{_{\!\perp}}\!)$], 
this implies
\begin{equation}
\delta b_{_{\!\ell}} \simeq  \ell^{\sigma_{_{\!b}}}, \quad \delta u_{_{\!\ell}} \simeq \ell^{\sigma_{_{\!u}}},
\quad \sigma_{_{\!b}} \sim 0.85, \quad \sigma_{_{\!u}} \sim -0.2, \label{eq:sig}
\end{equation}
i.e., while magnetic energy decreases at small scales kinetic energy increases
as it is expected to do for magnetic reconnection~\cite{Lazarian:1999p1657} where velocity 
is concentrated at small-scale current sheets~\cite{Rappazzo:2008p344}.

\begin{figure}
	\includegraphics[scale=.46]{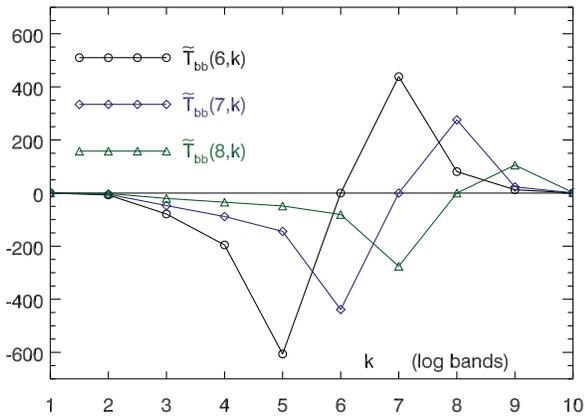}
	\caption{\label{fig3} (Color online) Energy flows  between scales (log bands)
	of the magnetic field. Log bands $(p_k]$=$(2^{k-2},2^{k-1}]$; see text.}
\end{figure}

Energy fluxes are shown in Fig.~\ref{fig2}. As their behavior is similar along the 
whole inertial range, here we plot the fluxes in and out of shell $Q=20$.
With respect to the equipartition case~(EQPT) \cite{Alexakis:2005p2633}, 
the most striking difference is the small 
value of the transfers between kinetic energy shells $\mathcal{T}_{_{uu}}^{(Q=20,K)}$,  
negligible with respect to the others. 
Indeed, the velocity eddies are not distorted by other velocity eddies as they are too
weak compared to the strength of the magnetic field: both the orthogonal component
$\mathbf{b}$ and the strong axial field $B_0$ are responsible for shaping the velocity 
field.

We analyze first the energy flows between shells of the magnetic field
$\mathcal{T}_{_{bb}}^{(Q=20,K)}$. They are negative for all K$<$Q and positive for all K$>$Q 
meaning that the field is receiving energy from shells at smaller K and giving energy to shells 
of greater K. In contrast to \textit{EQPT} for K$<$Q there is an almost constant small 
contribution from smaller K shells.
As in~\cite{Alexakis:2005p2633} a  similar ``tail'' is also present for 
$\mathcal{T}_{_{bu}}^{(Q=20,K)}$, the magnetic field
in shell 20 is receiving energy from smaller K shells of 
the velocity field and transferring it to larger K shells. 
$\mathcal{T}_{_{ub}}^{(Q=20,K)}= - \mathcal{T}_{_{bu}}^{(K,Q=20)}$ 
has the corresponding behavior.

The large peaks at K=20 represent the conversion of magnetic
to kinetic energy in the same shell and are due to 
$\mathcal{A}_{_{bu}}^{(K=20)}\! > 0$ 
($\mathcal{A}_{_{ub}}^{(K)}\! =\! -\mathcal{A}_{_{bu}}^{(K)}$).
Its large value is linked to the field-line tension of the dominant axial field 
$B_0$ as $\mathcal{A}_{_{bu}}$ is obtained from the linear terms 
$\propto c_{_A} \partial_z$ in eqs.~(\ref{eq:adim}), 
this is the Alfv\'en propagation term that in presence of a $b_{_{\!\perp}}$ contributes
with a velocity $u_{_{\!\perp}}$ of the same shape.
In Fig.~\ref{fig2} the values of the cross-field fluxes 
$\mathcal{T}_{_{bu}}$ and $\mathcal{T}_{_{ub}}$ for K=20 
without these contributions are shown  
with a diamond and a solid circle, respectively

\begin{figure}
	\includegraphics[scale=.46]{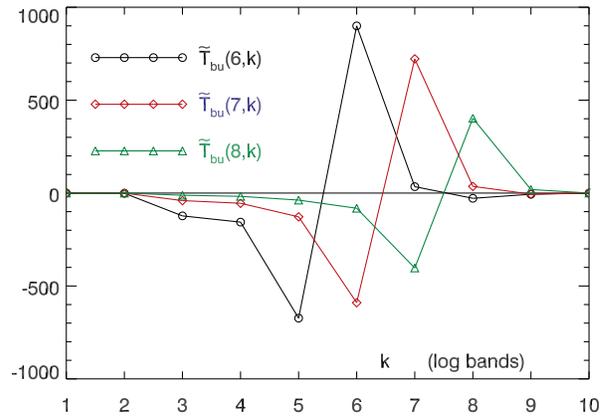}
	\caption{\label{fig4} (Color online) Energy fluxes from magnetic to velocity
	fields scales. Log bands, e.g., $(p_3]$=[3, 4] and $(p_6]$=[17, 32]}.
\end{figure}

The ``tails'' shown in Fig.~\ref{fig2} are also present for higher values of $Q$.
In~\cite{Alexakis:2005p2633}, they were present only in cross-fields transfers,
while here a comparable tail appears also in $\mathcal{T}_{_{bb}}$.
The tail in this work is due to the steeper spectrum of the magnetic field; 
this induces a higher spectral transfer at low $K$ because of the higher
value of $\mathbf{b}_{_K}$ in flux $\mathcal{T}_{_{bb}}$~[Eq.~(\ref{eq:t10})]
with respect to the EQPT case~\cite{Alexakis:2005p2633}.
This feature has been indicated as evidence of the nonlocal nature
of energy transfers in MHD turbulence~\cite{Alexakis:2005p2633}
because the cumulative transfers of farther shells seem to be more
important than those of closer shells.

However, in scaling models of MHD
turbulence~\cite{Goldreich:1995p1640,Galtier:2000p4461} and of  Parker's 
model~\cite{Rappazzo:2007p331,*Rappazzo:2008p344,*Rappazzo:2010p4099}
scales are defined as log bands of shells. 
In fact a single shell does not represent a \emph{scale} 
since, from the uncertainty principle, its associated field~[Eq.~(\ref{eq:shflt})]
is delocalized in space and cannot represent an \emph{eddy},
the building block of K41 phenomenology.

Thus in order to understand how
energy flows across scales we must use such log bands~\cite{Aluie:2010p4029}.
Log band $(p]$ is defined as the shells included in $(p/2, p]$, 
equally spaced on a logarithmic scale.
Considering $p_{n}\! = 2^{n-1}$ we will indicate these intervals with their index $n$: 
$(p_n] = ( 2^{n-2}, 2^{n-1}]$ (n=1,...,10). With a $1024^2$ grid we have 
``only'' $10$ distinct intervals.
Figures~\ref{fig3} and \ref{fig4} show the fluxes summed over these
log bands of shells, e.g., 
$\widetilde{T}_{_{bb}}^{(q,k)}\! =\! \mathcal{T}_{_{bb}}^{(\,(p_q], (p_k]\,)}$
 for q=6, 7 and 8.
The injection scale is now $(p_3]$=[3,4], while 
$(p_6]$=[17,32], $(p_7]$=[33,64], $(p_8]$=[65,128], etc..

Figures~\ref{fig3}-\ref{fig4} show that the apparently dominating contributions 
of distant shells (tails in Fig.~\ref{fig2}) strongly decrease when 
considering log bands.
In these bands the number of shells \emph{increases exponentially} at higher 
wave numbers and the aggregate effect of local transfers 
asymptotically dominates~\cite{Aluie:2010p4029,Aluie:2009p4035}.
Recently an analytical upper bound has been set for the locality of energy
transfers~\cite{Aluie:2010p4029}, although within these bounds
the energy transfers can be quite spread out~\cite{Beresnyak:2010p3363}.
In fact while cross energy transfers (Fig.~\ref{fig4}) are quite local
as energy flows between \emph{neighboring scales}  decrease swiftly,
the transfers between scales of the magnetic field (Fig.~\ref{fig3}) are more spread out.
However for the Parker problem we do not observe a direct flow
of energy between the forcing scale and the small scales~\cite{Yousef:2007p4206}.

\begin{figure}
	\begin{tikzpicture}[scale=4.2]
	   \draw [->]  (0, 0) node[left=2pt] {\large{$E_b$}} -- (1.5, 0)  node[right=2pt] {\large{$k$}};
	   \draw (1.6,-.2)  node { \large{ \textrm{(log bands)} } };
	   \draw [->]  (0,-.6) node[left=2pt] {\large{$E_u$}} -- (1.5,-.6)  node[right=2pt] {\large{$k$}};

	   \draw [thick] (intersection of .0,-.6--.25,0 and 0,-.16--2,-.16) coordinate (t) -- 
         	                         (intersection of t--.25,0 and 0,-.04--2,-.04);
	   \draw [->,thick] (intersection of .0,-.6--.25, 0 and 0,-.3--2,-.3) coordinate (t) -- 
       	                              (intersection of t--.25, 0 and 0,-.15--2,-.15);
 
	   \draw [thick] (.25, -.38) -- (.25, -.56);
	   \draw [->,thick]  (.25, -.04) -- (.25, -.4);

	   \draw [thick] (intersection of .25,-.6--.5,0 and 0,-.21--2,-.21) coordinate (t) -- 
       	                         (intersection of t--.5,0 and 0,-.04--2,-.04);
	   \draw [->,thick] (intersection of .25,-.6--.5, 0 and 0,-.56--2,-.56) coordinate (t) -- 
        	                              (intersection of t--.5, 0 and 0,-.2--2,-.2);

	   \draw [thick]        (.5, -.38) -- (.5, -.56);
	   \draw [->,thick]  (.5, -.04) -- (.5, -.4);

	   \draw [thick] (intersection of .5,-.6--.75,0 and 0,-.21--2,-.21) coordinate (t) -- 
         	                         (intersection of t--.75,0 and 0,-.04--2,-.04);
	   \draw [->,thick] (intersection of .5,-.6--.75, 0 and 0,-.56--2,-.56) coordinate (t) -- 
     	                              (intersection of t--.75, 0 and 0,-.2--2,-.2);

	   \draw [thick]        (.75, -.38) -- (.75, -.56);
	   \draw [->,thick]  (.75, -.04) -- (.75, -.4);

	   \draw [thick] (intersection of .75,-.6--1.,0 and 0,-.21--2,-.21) coordinate (t) -- 
         	                         (intersection of t--1.,0 and 0,-.04--2,-.04);
	   \draw [->,thick] (intersection of .75,-.6--1., 0 and 0,-.56--2,-.56) coordinate (t) -- 
        	                              (intersection of t--1., 0 and 0,-.2--2,-.2);

	  \draw [thick]        (1., -.38) -- (1., -.56);
	   \draw [->,thick]  (1., -.04) -- (1., -.4);

	   \draw [thick]  (intersection of 1.,-.6--1.25,0 and 0,-.21--2,-.21) coordinate (t) -- 
         	                         (intersection of t--1.25,0 and 0,-.04--2,-.04);
	   \draw [->,thick] (intersection of 1.,-.6--1.25, 0 and 0,-.56--2,-.56) coordinate (t) -- 
         	                              (intersection of t--1.25, 0 and 0,-.2--2,-.2);

	   \draw [thick]        (1.25, -.38) -- (1.25, -.56);
	   \draw [->,thick]  (1.25, -.04) -- (1.25, -.4);

	   \draw [thick] (intersection of 1.25,-.6--1.5,0 and 0,-.41--2,-.41) coordinate (t) -- 
         	                         (intersection of t--1.5,0 and 0,-.3--2,-.3);
	   \draw [->,thick] (intersection of 1.25,-.6--1.5, 0 and 0,-.56--2,-.56) coordinate (t) -- 
	   			 (intersection of t--1.5, 0 and 0,-.4--2,-.4);
	   \foreach \k in {0.25,0.5,...,1.25}
  		\filldraw [black] (\k,0) circle (.3pt);

		\draw [thick]   (.125,0) + (70:.125) arc (70:0:.125); 
		\draw [->,thick] (.125,0) + (90:.125) arc (90:60:.125); 

	   \foreach \k in {0.25,0.5,...,1.}
         		\draw [thick] (.125+\k,0) + (100:.125) arc (100:0:.125); 
            \foreach \k in {0.25,0.5,...,1.}
         		\draw [->,thick] (\k,0) arc (180:90:.125); 

            \draw [->,thick] (1.25,0) arc (180:120:.125); 
    	   \draw [thick]      (.125+1.25,0) + (130:.125) arc (130:90:.125); 

            \foreach \k in {0.25,0.5,...,1.25}
         	   \filldraw [black] (\k,-.6) circle (.3pt);
 
	   \draw [thick,dotted] (.125,-.6) + (300:.125) arc (300:360:.125); 
	   \draw [->,thick,dotted] (.125,-.6) + (270:.125) arc (270:300:.125); 
 
	   \foreach \k in {0.25,0.5,...,1.}
            \draw [thick,dotted] (.125+\k,-.6) + (270:.125) arc (270:360:.125); 
	   \foreach \k in {0.25,.5,...,1.}
         	   \draw [->,thick,dotted] (\k,-.6) arc (180:270:.125); 

	   \draw [->,thick,dotted] (1.25,-.6) arc (180:240:.125); 
 	   \draw [thick,dotted] (.125+1.25,-.6) + (240:.125) arc (240:270:.125); 
	\end{tikzpicture}
	\caption{\label{fig5} Diagram of energy flows between fields and scales.}
\end{figure}
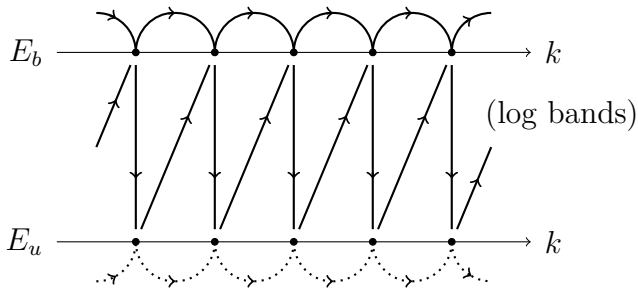

Overall energy is transferred from larger to smaller scales in a similar fashion as in MHD
turbulence with energies in equipartition, except for  
the velocity field that is too weak compared to the magnetic field.
As a result the magnetic field creates and shapes the velocity field.
In fact from cross-field flows (Fig.~\ref{fig4}) we see that magnetic field line tension enhanced 
by line tying, and predominantly represented by the fluxes $\mathcal{A}_{_{bu}}$,
converts magnetic energy to kinetic energy \emph{at the same scale}. 
In turn kinetic energy at larger scales is converted to magnetic energy at smaller scales, 
due to the magnetic stretching term.
The magnetic advection term transfers a (smaller) fraction of
energy toward smaller magnetic field scales (Fig.~\ref{fig3}).
A pictorial summary of the cascade is shown in Fig.~\ref{fig5}
(magnetic flux spread not shown);
the repeated conversion of kinetic to magnetic energies by
the cross-field flows effectively cascades energy toward the 
small scales.

The upper bound for the locality of energy transfers found by~\cite{Aluie:2010p4029}
has been restricted to the case $ 0 < \sigma_{_{\!b,u}} < 1$ in analogy to the 
hydrodynamic case~\cite{Aluie:2009p4035}
but this condition is overrestrictive for MHD  and we show that those 
bounds are valid also for the case [Eq.~(\ref{eq:sig})] where $\sigma_{_{\!b}} \sim 0.85$,
but  $\sigma_{_{\!u}} \sim -0.2$ instead of the standard 
$\sigma_{_{\!b,u}} \sim 1/3$.

If $[K]$ is the log band of shells $[K/2,K]$, heuristic scalings [Eq.~(\ref{eq:sig})]
can be written more precisely for the generic band-summed field 
$\mathbf{a}^{[K]} = \sum \mathbf{a}_{_K}$ with $K \in [K]$  [Eq.~(\ref{eq:shflt})] as
\begin{equation}
\hspace{-.001em}
\langle | \mathbf{a}^{[K]} |^3 \rangle^{1/3} \simeq K^{-\sigma_{_{\!a}}}
\quad \rightarrow \quad
\langle | \nabla \mathbf{a}^{[K]} |^3 \rangle^{1/3} \simeq K^{1-\sigma_{_{\!a}}}.
\hspace{-.5em}
\end{equation}
Since the scaling for the derivative is valid independently of the sign of $\sigma_{_{\!a}}$,
following~\cite{Aluie:2010p4029} the bound for the locality of
cross-field transfers between band-summed fields is{\setlength\arraycolsep{-1em}
\begin{eqnarray}
&&\left|\, T_{_{bu}}^{(\, [Q],[K]\, )}\, \right| \le 
K^{-(\sigma_u +\sigma_b)}\, Q^{1- \sigma_b}, \quad \textrm{for} \quad Q \ll K \quad \label{ae1} \\[.1em]
&&\left|\, T_{_{bu}}^{(\, [Q],[K]\, )}\, \right| \le 
K^{1 - \sigma_u}\, Q^{-2 \sigma_b}, \qquad \  \ \textrm{for} \quad Q \gg K \label{ae2}
\end{eqnarray}
}At fixed $K$, contributions from smaller bands  [Eq.~(\ref{ae1})] are negligible if
$\sigma_{_{\!b}} < 1$ so as from bigger bands [Eq.~(\ref{ae2})] for $\sigma_{_{\!b}} > 0$.

In similar fashion, we obtain the following bound for magnetic fluxes:
\begin{equation}
\left|\, T_{_{bb}}^{(\, [Q],[K]\, )}\, \right| \le 
K^{-(\sigma_u +\sigma_b)}\, Q^{1- \sigma_b}, \quad \textrm{for} \quad  Q \ll K  \label{rv1}
\end{equation}

\begin{equation}
\left|\, T_{_{bb}}^{(\, [Q],[K]\, )}\, \right| \le 
K^{1 - \sigma_b}\, Q^{-( \sigma_u + \sigma_b)}, \quad \textrm{for} \quad  Q \gg K \label{rv2}
\end{equation}
The requirement for asymptotic locality is still $\sigma_{_{\!b}} < 1$ for Eq.~(\ref{rv1}), but
$\sigma_{_{\!u}} + \sigma_{_{\!b}} > 0$ for  Eq.~(\ref{rv2}), all satisfied
in our case.

The sign of the exponents containing $\sigma_{_{\!u}}$ remains unaltered
with respect to the classic case $\sigma_{_{\!b,u}} \sim 1/3$,
but while in Eqs.~(\ref{ae1}) and (\ref{rv1}) $K^{-0.65}$ decreases for large $K$ in 
Eqs.~(\ref{ae2}) and (\ref{rv2}) we have a positive power 
($K^{1.2}$ and $K^{0.15}$, respectively)  as in the standard case~\cite{Aluie:2010p4029}
with $K^{2/3}$. As in the hydrodynamic case~\cite{Aluie:2009p4035} the
origin of this pathological behavior stems from the use of H\"older inequality
for fluxes [Eqs.~(\ref{eq:t9})-(\ref{eq:t10})] to set the upper bound, since only the absolute
values of their terms are considered, and any cancellation effect due to the scalar 
products in Eqs.~(\ref{eq:t9})-(\ref{eq:t10}) is neglected.
This conclusion is reinforced by the fact that for Parker problem fluxes, Eqs.~(\ref{eq:t9})-(\ref{eq:t10})
exhibit scalings (not shown) in $Q$ well below upper bounds [Eqs.~(\ref{ae1})-(\ref{rv2})].

\begin{acknowledgments}
Research supported in part by the Jet Propulsion Laboratory, California 
Institute of Technology under a contract with NASA, and in part
by the European Commission through the SOLAIRE Network (MRTN-CT-2006-035484) 
and by the Spanish Ministry of Research and Innovation through projects
 AYA2007-66502 and CSD2007-00050, and by 
NSF grants AGS-1063439
and (SHINE) ATM-0752135
as well as NASA Heliophysics Theory Program
grant ATM-0752135.
Simulations  were carried out through NASA Advanced Supercomputing 
SMD Award Nos.\ 09-1112 and 10-1633, and a Key Project at CINECA.
\end{acknowledgments}


\begin{thebibliography}{38}%
\makeatletter
\providecommand \@ifxundefined [1]{%
 \@ifx{#1\undefined}
}%
\providecommand \@ifnum [1]{%
 \ifnum #1\expandafter \@firstoftwo
 \else \expandafter \@secondoftwo
 \fi
}%
\providecommand \@ifx [1]{%
 \ifx #1\expandafter \@firstoftwo
 \else \expandafter \@secondoftwo
 \fi
}%
\providecommand \natexlab [1]{#1}%
\providecommand \enquote  [1]{``#1''}%
\providecommand \bibnamefont  [1]{#1}%
\providecommand \bibfnamefont [1]{#1}%
\providecommand \citenamefont [1]{#1}%
\providecommand \href@noop [0]{\@secondoftwo}%
\providecommand \href [0]{\begingroup \@sanitize@url \@href}%
\providecommand \@href[1]{\@@startlink{#1}\@@href}%
\providecommand \@@href[1]{\endgroup#1\@@endlink}%
\providecommand \@sanitize@url [0]{\catcode `\\12\catcode `\$12\catcode
  `\&12\catcode `\#12\catcode `\^12\catcode `\_12\catcode `\%12\relax}%
\providecommand \@@startlink[1]{}%
\providecommand \@@endlink[0]{}%
\providecommand \url  [0]{\begingroup\@sanitize@url \@url }%
\providecommand \@url [1]{\endgroup\@href {#1}{\urlprefix }}%
\providecommand \urlprefix  [0]{URL }%
\providecommand \Eprint [0]{\href }%
\providecommand \doibase [0]{http://dx.doi.org/}%
\providecommand \selectlanguage [0]{\@gobble}%
\providecommand \bibinfo  [0]{\@secondoftwo}%
\providecommand \bibfield  [0]{\@secondoftwo}%
\providecommand \translation [1]{[#1]}%
\providecommand \BibitemOpen [0]{}%
\providecommand \bibitemStop [0]{}%
\providecommand \bibitemNoStop [0]{.\EOS\space}%
\providecommand \EOS [0]{\spacefactor3000\relax}%
\providecommand \BibitemShut  [1]{\csname bibitem#1\endcsname}%
\let\auto@bib@innerbib\@empty

\bibitem[Reale (2010)]{re10}
  \BibitemOpen
   F.~Reale, Living Rev. Solar Phys. \textbf{7}, (2010), 5 \BibitemShut {NoStop}%
\bibitem [{\citenamefont {Withbroe}\ and\ \citenamefont
  {Noyes}(1977)}]{Withbroe:1977p363}%
  \BibitemOpen
  \bibfield  {author} {\bibinfo {author} {\bibfnamefont {G.~L.}\ \bibnamefont
  {Withbroe}}\ and\ \bibinfo {author} {\bibfnamefont {R.~W.}\ \bibnamefont
  {Noyes}},\ }\href@noop {} {\bibfield  {journal} {\bibinfo  {journal} {Annu.
  Rev. Astron. Astrophys.}\ }\textbf {\bibinfo {volume} {15}},\ \bibinfo
  {pages} {363} (\bibinfo {year} {1977})}\BibitemShut {NoStop}%
\bibitem [{\citenamefont {Parker}(1972)}]{Parker:1972p1352}%
  \BibitemOpen
  \bibfield  {author} {\bibinfo {author} {\bibfnamefont {E.~N.}\ \bibnamefont
  {Parker}},\ }\href@noop {} {\bibfield  {journal} {\bibinfo  {journal}
  {Astrophys. J.}\ }\textbf {\bibinfo {volume} {174}},\ \bibinfo {pages} {499}
  (\bibinfo {year} {1972})}\BibitemShut {NoStop}%
\bibitem [{\citenamefont {Einaudi}\ \emph {et~al.}(1996)}]{Einaudi:1996p729}%
  \BibitemOpen
  \bibfield  {author} {\bibinfo {author} {\bibfnamefont {G.}~\bibnamefont
  {Einaudi}}\ \emph {et~al.},\ }\href@noop
  {} {\bibfield  {journal} {\bibinfo  {journal} {Astrophys. J. Lett.}\ }\textbf
  {\bibinfo {volume} {457}},\ \bibinfo {pages} {L113} (\bibinfo {year}
  {1996})}\BibitemShut {NoStop}%
\bibitem [{\citenamefont {Dmitruk}\ and\ \citenamefont
  {G\'omez}(1997)}]{Dmitruk:1997p1563}%
  \BibitemOpen
  \bibfield  {author} {\bibinfo {author} {\bibfnamefont {P.}~\bibnamefont
  {Dmitruk}}\ and\ \bibinfo {author} {\bibfnamefont {D.~O.}\ \bibnamefont
  {G\'omez}},\ }\href@noop {} {\bibfield  {journal} {\bibinfo  {journal}
  {Astrophys. J. Lett.}\ }\textbf {\bibinfo {volume} {484}},\ \bibinfo {pages}
  {L83} (\bibinfo {year} {1997})}\BibitemShut {NoStop}%
\bibitem [{\citenamefont {Einaudi}\ and\ \citenamefont
  {Velli}(1999)}]{Einaudi:1999p707}%
  \BibitemOpen
  \bibfield  {author} {\bibinfo {author} {\bibfnamefont {G.}~\bibnamefont
  {Einaudi}}\ and\ \bibinfo {author} {\bibfnamefont {M.}~\bibnamefont
  {Velli}},\ }\href@noop {} {\bibfield  {journal} {\bibinfo  {journal} {Phys.
  Plasmas}\ }\textbf {\bibinfo {volume} {6}},\ \bibinfo {pages} {4146}
  (\bibinfo {year} {1999})}\BibitemShut {NoStop}%
\bibitem [{\citenamefont {Dmitruk}\ \emph {et~al.}(2003)\citenamefont
  {Dmitruk}, \citenamefont {G{\'o}mez},\ and\ \citenamefont
  {Matthaeus}}]{Dmitruk:2003p499}%
  \BibitemOpen
  \bibfield  {author} {\bibinfo {author} {\bibfnamefont {P.}~\bibnamefont
  {Dmitruk}}, \bibinfo {author} {\bibfnamefont {D.~O.}\ \bibnamefont
  {G{\'o}mez}}, \ and\ \bibinfo {author} {\bibfnamefont {W.~H.}\ \bibnamefont
  {Matthaeus}},\ }\href@noop {} {\bibfield  {journal} {\bibinfo  {journal}
  {Phys. Plasmas}\ }\textbf {\bibinfo {volume} {10}},\ \bibinfo {pages} {3584}
  (\bibinfo {year} {2003})}\BibitemShut {NoStop}%
\bibitem [{\citenamefont {Rappazzo}\ \emph {et~al.}(2007)}]{Rappazzo:2007p331}%
  \BibitemOpen
  \bibfield  {author} {\bibinfo {author} {\bibfnamefont {A.~F.}\ \bibnamefont
  {Rappazzo}}\ \emph {et~al.},\
  }\href@noop {} {\bibfield  {journal} {\bibinfo  {journal} {Astrophys. J.}\
  }\textbf {\bibinfo {volume} {657}},\ \bibinfo {pages} {L47} (\bibinfo {year}
  {2007})}\BibitemShut {NoStop}%
\bibitem [{\citenamefont {Rappazzo}\ \emph {et~al.}(2008)}]{Rappazzo:2008p344}%
  \BibitemOpen
  \bibfield  {author} {\bibinfo {author} {\bibfnamefont {A.~F.}\ \bibnamefont
  {Rappazzo}}\ \emph {et~al.},\
  }\href@noop {} {\bibfield  {journal} {\bibinfo  {journal} {Astrophys. J.}\
  }\textbf {\bibinfo {volume} {677}},\ \bibinfo {pages} {1348} (\bibinfo {year}
  {2008})}\BibitemShut {NoStop}%
\bibitem [{\citenamefont {Rappazzo}\ \emph {et~al.}(2010)}]{Rappazzo:2010p4099}%
  \BibitemOpen
  \bibfield  {author} {\bibinfo {author} {\bibfnamefont {A.~F.}\ \bibnamefont
  {Rappazzo}}, \bibinfo {author} {\bibfnamefont {M.}~\bibnamefont {Velli}}, \
  and\ \bibinfo {author} {\bibfnamefont {G.}~\bibnamefont {Einaudi}},\
  }\href@noop {} {\bibfield  {journal} {\bibinfo  {journal} {Astrophys. J.}\
  }\textbf {\bibinfo {volume} {722}},\ \bibinfo {pages} {65} (\bibinfo {year}
  {2010})}\BibitemShut {NoStop}%
\bibitem [Galtier \emph{et~al.}(2000)]{Galtier:2000p4461}%
  S.~Galtier \emph{et~al.}, J. Plasma Phys. \textbf{63}, 447 (2000). 
\bibitem [{\citenamefont {Goldreich}\ and\ \citenamefont
  {Sridhar}(1995)}]{Goldreich:1995p1640}%
  \BibitemOpen
  \bibfield  {author} {\bibinfo {author} {\bibfnamefont {P.}~\bibnamefont
  {Goldreich}}\ and\ \bibinfo {author} {\bibfnamefont {S.}~\bibnamefont
  {Sridhar}},\ }\href@noop {} {\bibfield  {journal} {\bibinfo  {journal}
  {Astrophys. J.}\ }\textbf {\bibinfo {volume} {438}},\ \bibinfo {pages} {763}
  (\bibinfo {year} {1995})}\BibitemShut {NoStop}%
\bibitem [Lazarian \& Vishniac(1999)]{Lazarian:1999p1657}%
  A.~Lazarian and E.~T.~Vishniac, Astrophys. J. \textbf{517}, 700  (1999).
\bibitem [Dar \emph{et~al.}(2001)]{Dar:2001p2747}%
  G.~Dar, M.~K.~Verma, and V.~Eswaran, Physica D \textbf{157}, 207 (2001).
\bibitem [{\citenamefont {Alexakis}\ \emph {et~al.}(2005)\citenamefont
  {Alexakis}, \citenamefont {Mininni},\ and\ \citenamefont
  {Pouquet}}]{Alexakis:2005p2633}%
  \BibitemOpen
  \bibfield  {author} {\bibinfo {author} {\bibfnamefont {A.}~\bibnamefont
  {Alexakis}}, \bibinfo {author} {\bibfnamefont {P.~D.}\ \bibnamefont
  {Mininni}}, \ and\ \bibinfo {author} {\bibfnamefont {A.}~\bibnamefont
  {Pouquet}},\ }\href@noop {} {\bibfield  {journal} {\bibinfo  {journal} {Phys.
  Rev. E}\ }\textbf {\bibinfo {volume} {72}},\ \bibinfo {pages} {46301}
  (\bibinfo {year} {2005})}\BibitemShut {NoStop}%
\bibitem [{\citenamefont {Aluie}\ and\ \citenamefont
  {Eyink}(2010)}]{Aluie:2010p4029}%
  \BibitemOpen
  \bibfield  {author} {\bibinfo {author} {\bibfnamefont {H.}~\bibnamefont
  {Aluie}}\ and\ \bibinfo {author} {\bibfnamefont {G.~L.}\ \bibnamefont
  {Eyink}},\ }\href@noop {} {\bibfield  {journal} {\bibinfo  {journal} {Phys.
  Rev. Lett.}\ }\textbf {\bibinfo {volume} {104}},\ \bibinfo {pages} {81101}
  (\bibinfo {year} {2010})}\BibitemShut {NoStop}%
\bibitem [{\citenamefont {Kadomtsev}\ and\ \citenamefont
  {Pogutse}(1974)}]{Kadomtsev:1974p283}%
  \BibitemOpen
  \bibfield  {author} {\bibinfo {author} {\bibfnamefont {B.~B.}\ \bibnamefont
  {Kadomtsev}}\ and\ \bibinfo {author} {\bibfnamefont {O.~P.}\ \bibnamefont
  {Pogutse}},\ }\href@noop {} {\bibfield  {journal} {\bibinfo  {journal} {Sov.
  Phys. JETP}\ }\textbf {\bibinfo {volume} {38}},\ \bibinfo {pages} {283}
  (\bibinfo {year} {1974})}\BibitemShut {NoStop}%
\bibitem [{\citenamefont {Strauss}(1976)}]{Strauss:1976p1438}%
  \BibitemOpen
  \bibfield  {author} {\bibinfo {author} {\bibfnamefont {H.~R.}\ \bibnamefont
  {Strauss}},\ }\href@noop {} {\bibfield  {journal} {\bibinfo  {journal} {Phys.
  Fluids}\ }\textbf {\bibinfo {volume} {19}},\ \bibinfo {pages} {134} (\bibinfo
  {year} {1976})}\BibitemShut {NoStop}%
\bibitem [Aluie \& Eyink(2009)]{Aluie:2009p4035}%
  H.~Aluie  and G.~Eyink, Phys. Fluids \textbf{21}, 115108 (2009).%
\bibitem [Beresnyak \& Lazarian(2010)]{Beresnyak:2010p3363}%
  A.~Beresnyak and A.~Lazarian, Astrophys. J. \textbf{722}, L110  (2010).%
\bibitem [{\citenamefont {Yousef}\ \emph {et~al.}(2007)\citenamefont {Yousef},%
  \citenamefont {Rincon},\ and\ \citenamefont
  {Schekochihin}}]{Yousef:2007p4206}%
  \BibitemOpen
  \bibfield  {author} {\bibinfo {author} {\bibfnamefont {T.~A.}\ \bibnamefont
  {Yousef}}, \bibinfo {author} {\bibfnamefont {F.}~\bibnamefont {Rincon}}, \
  and\ \bibinfo {author} {\bibfnamefont {A.~A.}\ \bibnamefont {Schekochihin}},\
  }\href@noop {} {\bibfield  {journal} {\bibinfo  {journal} {J. Fluid Mech.}\
  }\textbf {\bibinfo {volume} {575}},\ \bibinfo {pages} {111} (\bibinfo {year}
  {2007})}\BibitemShut {NoStop}%
\end{thebibliography}
\end{document}